\documentclass[conference]{IEEEtran}
\IEEEoverridecommandlockouts

\usepackage{cite}
\usepackage{amsmath,amssymb,amsfonts}
\usepackage{algorithm}
\usepackage{algorithmic}
\usepackage{graphicx}
\usepackage{textcomp}
\usepackage{xcolor}
\def\BibTeX{{\rm B\kern-.05em{\sc i\kern-.025em b}\kern-.08em
    T\kern-.1667em\lower.7ex\hbox{E}\kern-.125emX}}

\usepackage{wrapfig}

\usepackage{enumitem}
\usepackage{mathtools}
\usepackage{listings}
\usepackage{subfig}
\usepackage{color}
      \usepackage{graphicx}
      \usepackage{caption}
      \usepackage{subcaption}
      \usepackage{comment}
      \usepackage{tabularx}
      \usepackage{float}
      \usepackage[export]{adjustbox}
    \usepackage{setspace}

\usepackage{multirow}

\newtheorem{definition}{\textbf{\noindent Definition}}
\newtheorem{theorem}{\textbf{Theorem}}[section]
\newtheorem{lemma}[theorem]{\textbf{Lemma}}

\begin{document}

\title{Scalable Substructure Discovery Algorithm For Homogeneous Multilayer Networks}

\author{\IEEEauthorblockN{Arshdeep Singh, Abhishek Santra and Sharma Chakravarthy }
\IEEEauthorblockA{
\textit{IT Lab and CSE Department, University of Texas at Arlington}\\
axs9120@mavs.uta.edu, abhishek.santra@mavs.uta.edu, sharmac@cse.uta.edu }
}

\maketitle

\begin{abstract}
Graph mining analyzes real-world graphs to find core substructures (connected subgraphs) in applications modeled as graphs. Substructure discovery is a process that involves identifying meaningful patterns, structures, or components within a large data set. These substructures can be of various types, such as frequent patterns, motifs, or other relevant features within the data. 

To model complex data sets -- with multiple types of entities and relationships -- multilayer networks (or MLNs) have been shown to be more effective as compared to simple and attributed graphs. Analysis algorithms on MLNs using the decoupling approach have been shown to be both efficient and accurate. Hence, this paper focuses on substructure discovery in \textbf{homogeneous multilayer networks} (one type of MLN) using a novel decoupling-based approach. In this approach, each layer is processed \textbf{independently}, and then the results from two or more layers are \textbf{composed} to identify substructures in the entire MLN. The algorithm is designed and implemented, including the composition part, using one of the distributed processing frameworks (the Map/Reduce paradigm) to provide scalability. 

After establishing the correctness, we analyze the speedup and response time of the proposed algorithm and approach through extensive experimental analysis on large synthetic and real-world data sets with diverse graph characteristics.

\end{abstract}

\begin{IEEEkeywords}
Homogeneous Multilayer Networks, Substructure Discovery, Decoupling approach, Accuracy \& Scalability
\end{IEEEkeywords}

\section{Introduction}
\label{section:introduction}

\noindent Substructure discovery is an unsupervised approach to data mining to infer new knowledge and is typically applied to graph representations. Frequent patterns are a common concept in data mining that refers to item sets, subsequences, or substructures that occur frequently in a given data set, meeting or exceeding a user-defined threshold. 
By identifying frequent subgraphs, we can gain insights into the underlying structure of the data and identify recurring patterns or motifs.

Graph models have been used to analyze the World Wide Web's structure \cite{analysis_graph21}, 
social-media data, bio-informatics data~\cite{bijung_graph1946}, atoms and covalent relationships in chemistry \cite{doi:10.1021/ci00047a033}, etc. Graphs are better than other data representations that embed inherent relationships among objects or entities and are also easy to understand. They use vertices and edges where each vertex of the graph will correspond to an entity and each edge is a relationship between two entities. Applications can be modeled using several graph alternatives: (i) Simple graphs, (ii) Attributed graphs, and (iii) Multilayer networks.

\noindent \textbf{Need for Multilayer Networks: } Social networks and many other applications often include diverse types of nodes and relationships between them. Modeling or representing them as simple graphs (also termed a network) by using only one type of node and relationship often leads to an oversimplification of reality and information loss. Attributed graphs are better, but lack algorithms for analysis. 

Sociologists have long acknowledged the significance of using distinct types of connectivity among the same group of individuals \cite{wasserman1994social}. Therefore, such systems are better studied through multilayer networks.

Multilayer networks are a more general framework that allows for the analysis of complex systems with multiple interacting units and inter-dependencies that are not properly captured by simple graphs. In a multilayer network, each layer represents a different type of interaction or relationship between the system's constituents. For example, in a social network, one layer may represent friendships between individuals, while another layer may represent professional LinkedIn connections. By analyzing the interactions between layers, it is possible to gain insights into the emergent behavior of the system as a whole. 

\begin{figure}[h]
\vspace{-11pt}
  \centering
  \includegraphics[keepaspectratio=true,width=0.85\columnwidth]{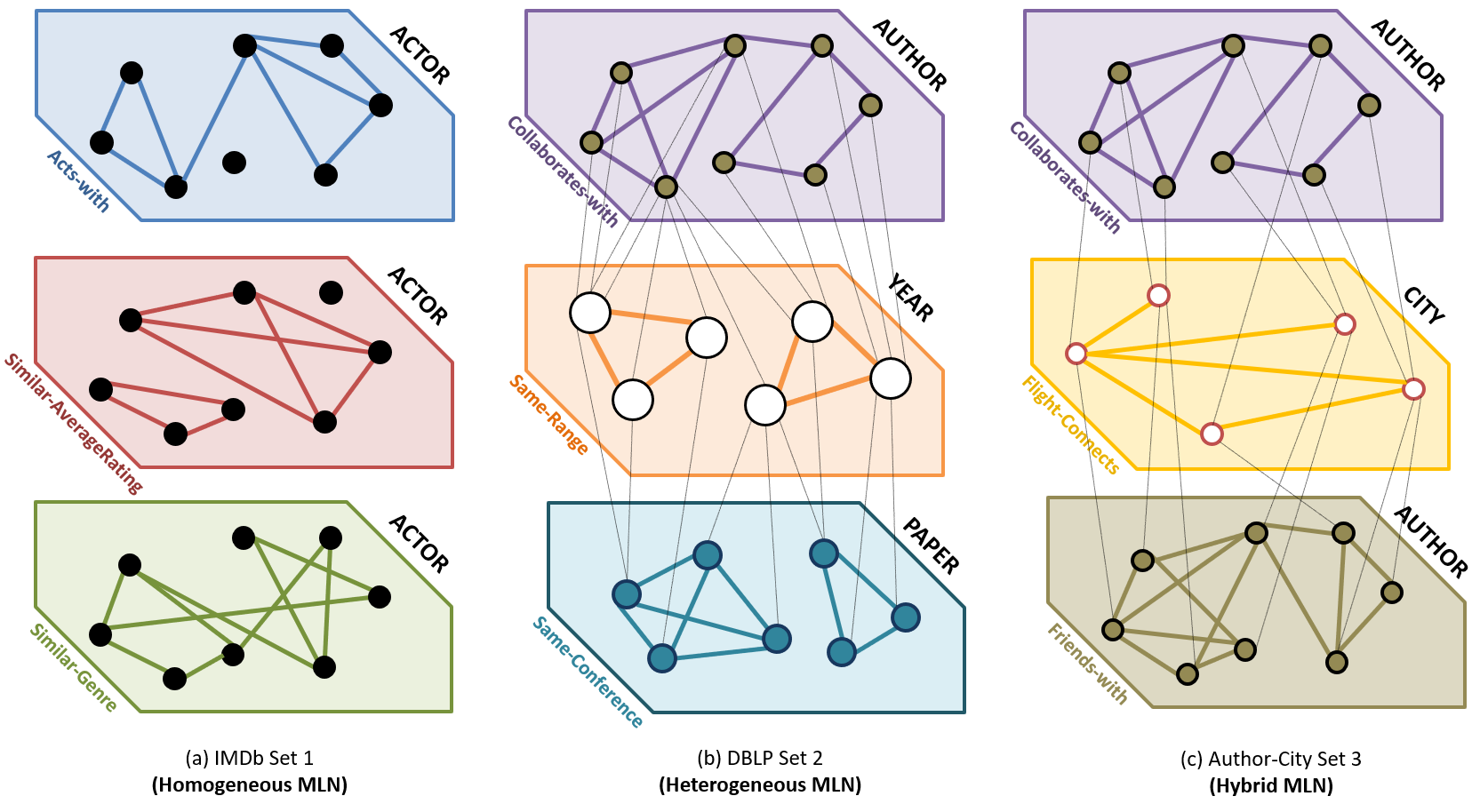}
  \caption{MLN Types}
  \label{fig:mln-types}
\vspace{-11pt}
\end{figure}

Multilayer networks can be of several types, including (i) homogeneous (HoMLNs, Figure~\ref{fig:mln-types}(a)), (ii) heterogeneous (HeMLNs, Figure~\ref{fig:mln-types}(b)), and (iii) hybrid (HyMLNs, Figure~\ref{fig:mln-types}(c)), depending on the type of layers and connections between them. Figure \ref{fig:mln-types} shows these types of multilayer networks. Since the focus of this paper is HoMLNs, we start with a formal definition of a MLN.

\begin{definition}
A multilayer network $MLN(G,X)$, is defined by two sets of graphs. The set $G=\{G_1, G_2, \ldots, G_n\}$ contains \textbf{simple} graphs (one for each layer), where $G_i (V_i, E_i)$ is defined by a set of vertices $V_i$ and a set of edges $E_i$. An edge $e(v,u) \in E_i$, connects vertices $v$ and $u$, where $v,u\in V_i$. The set $X=\{X_{1,2}, X_{1,3}, \ldots, X_{m-1,m}\}$ consists of \textbf{bipartite} graphs\footnote{The number of bipartite graphs can be less or even more than the number of graphs/layers. That is, $m$ can be $<=$ or even $>$ $n$.}. Each graph $X_{i,j} (V_i, V_j, L_{i,j})$ is defined by two sets of vertices $V_i$ and $V_j$ and a set of edges (or links) $L_{i,j}$, such that for every link $l(a,b) \in L_{i,j}$,  $a\in V_i$ and $b \in V_j$, where $V_i$ ($V_j$) is the vertex set of graph $G_i$ ($G_j$). Some of $X_{i,j}$ can be empty. \textbf{For HoMLNs, inter-layer graphs are implicit as nodes in layers form a common subset having same identity.}
\end{definition}

\noindent \textbf{Problem Addressed:} Given a homogeneous MLN (HoMLN) with $l$ number of layers --  \textit{G$_1$ (V, E$_1$), G$_2$ (V, E$_2$), ..., G$_l$ (V, E$_l$)}, where $V$ and $E_i$ are the vertex and edge set in the $i^{th}$ layer -- the goal is to discover \textit{interesting substructures} of the HoMLN for any $r$ layers \textbf{correctly and efficiently} using the decoupling approach. For correctness, $r$ HoMLN layers under consideration are conflated into a simple graph using the Boolean OR operator, and substructures discovered on that simple graph are used as the Ground Truth (GT).

\noindent \textbf{Why Map/Reduce?:} We have used the Map/Reduce paradigm as an example of the distributed and parallel processing approach. Without loss of generality, any other paradigm (e.g., Spark, Pregel) can be used in its stead without the need to modify the algorithms presented in this paper. Implementation will change slightly based on the choice of the framework.

\noindent \textit{Contributions of this paper are:}
\begin{itemize}
\item \textbf{Map/Reduce Algorithm} for substructure discovery in a layer.
\item \textbf{Map/Reduce Composition algorithm} to correctly generate substructures spanning  Homogeneous layers.
\item Map/Reduce \textbf{Partitioned approach} to layer graph processing.
\item \textbf{Correctness proof} of both independent expansion and the composition algorithm (Ho-ICA).
\item Extensive experimental analysis on a large number of \textbf{synthetic and real-world} graphs.
\item \textbf{Accuracy} with ground truth, response time, and speedup comparisons.
\end{itemize}

\noindent \textbf{Roadmap: } MLN analysis alternatives are briefly discussed in Section~\ref{sec:mln-analysis-alternatives} to highlight our choice. Section~\ref{sec:related-work} has related work. Preliminaries and terminology are outlined in Section~\ref{sec:preliminaries}.
Section~\ref{sec:architecture-algorithm} details the substructure and composition algorithm and Map/Reduce implementation of the proposed substructure discovery. Correctness of algorithms is proved in Section~\ref{sec:homln-correctness}. Detailed experimental evaluation is given in Section~\ref{sec:experiments-validation}. Section~\ref{sec:conclusion} has conclusions. 
\section{MLN Analysis Alternatives and Challenges}
\label{sec:mln-analysis-alternatives}
\begin{figure}[h]
   \vspace{-10pt}
   \centering \includegraphics[keepaspectratio=true,width=1\linewidth]{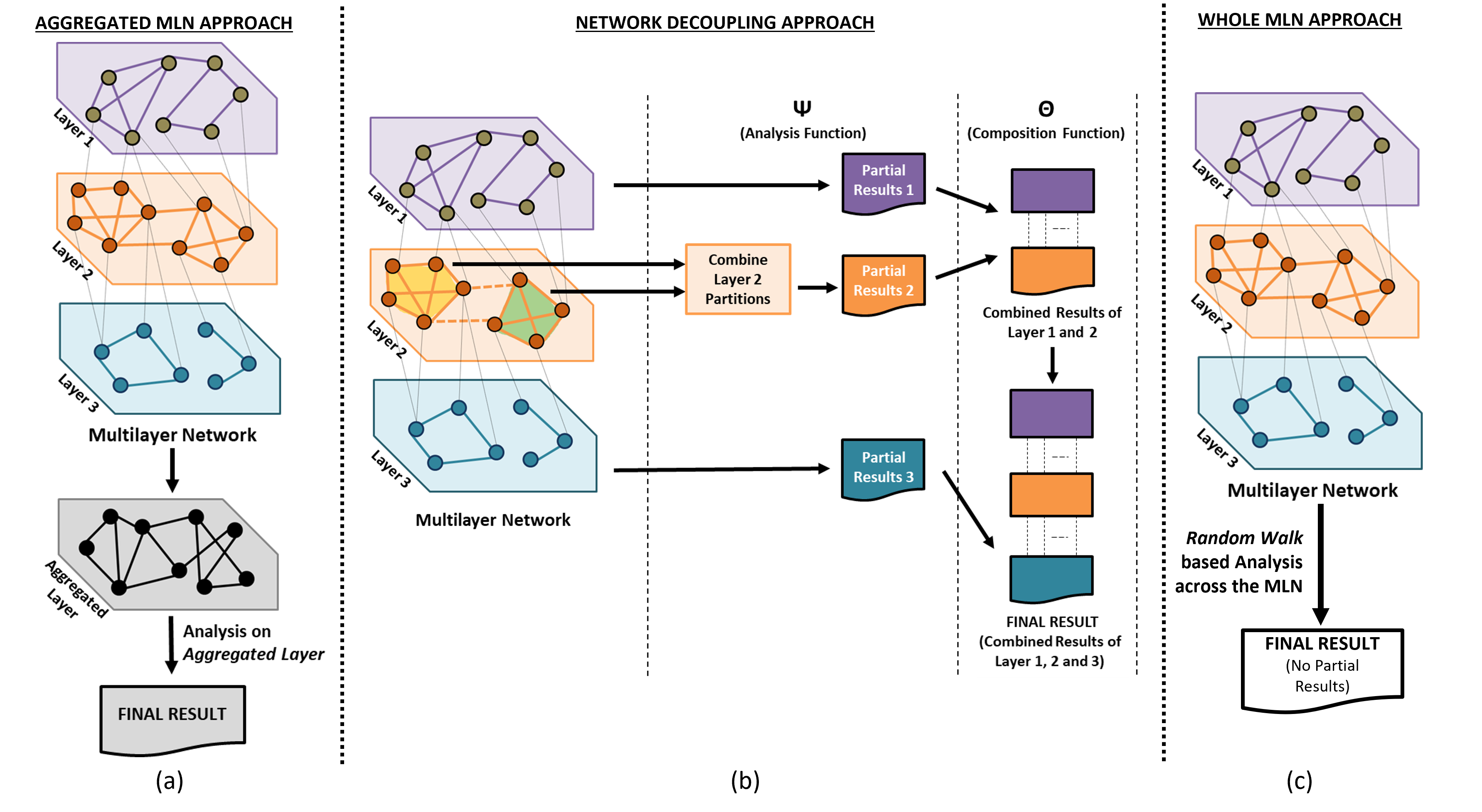}
   \caption{Lossy (a),  Decoupling  (b), and MLN (c) approaches}
   \label{fig:3-approaches}
   \vspace{-10pt}
\end{figure}

\noindent Figure~\ref{fig:3-approaches} shows three alternatives to perform analysis on MLNs. Figure~\ref{fig:3-approaches}(a) shows the traditional approach, where a MLN is conflated into a simple graph using type-independent~\cite{LayerAggDomenicoNAL14} or projection-based~\cite{berenstein2016multilayer} approaches. 
As both ignore type information for the transformation, they do not support the structure and semantic preservation of MLNs.
Moreover, as observed in the literature, \textit{without additional mappings}, these approaches are likely to result in some information loss and distortion of properties~\cite{MultiLayerSurveyKivelaABGMP13}, or hide the effect of different entity types and/or different intra- or inter-layer relationships~\cite{DeDomenico201318469}. At the other end of the spectrum, Figure~\ref{fig:3-approaches}(c) shows computing the result by traversing the whole MLN as a single graph. Although this has been implemented for community detection (e.g., Infomap recently, ~\cite{wilson2017community}), this is likely to be computationally expensive and not flexible as the number of layers and data sizes become large and new MLN algorithms need to be developed for each analysis. 

Figure~\ref{fig:3-approaches}(b), on the other hand, shows an approach (termed \textbf{network decoupling}) where the analysis metric for each layer is computed independently (possibly in parallel) during the analysis ($\Psi$) phase \textbf{only once} and  results composed using a binary operator $\Theta$ as shown in~\cite{DBLP:journals/fdata/SantraIMCB24, ICCS/SantraBC17,IC3K2023/PavelRSC23,BDS2023/MukundaRSC,DBLP:journals/dke/SantraKBC22}. Other layer or inter-layer information is \textbf{not used} while processing a layer! This approach has been shown to be effective for Boolean operations without aggregating and losing type information. Furthermore, it has been shown to be more efficient than the approaches shown in Figure~\ref{fig:3-approaches}(a) or ~Figure~\ref{fig:3-approaches}(c). 

The challenges in developing a substructure discovery algorithm are: \textit{i) enumerating all connected subgraphs of any size in a given graph or forest (completeness), ii) identifying duplicates, if any, and remove them (soundness), iii) counting isomorphic substructures to apply the metric for each distinct substructure and ranking them, and iv) retaining top-beam substructures for the next iteration.} The enumeration is typically done by iteratively increasing the connected subgraph size by 1. Evaluation of each substructure with the desired metric (Frequency or Minimum Description Length) is carried out after each iteration. To contain the search space, a heuristic is applied to carry top-\textit{beam} (\textit{beam} is a parameter) results from each iteration. Current algorithms do this on a simple graph or forest. \textbf{How the decoupling approach has been adapted to the HoMLN substructure discovery is the focus of this paper and is elaborated in Section~\ref{sec:architecture-algorithm}.}

\section{Related Work}
\label{sec:related-work}

\noindent SUBDUE was the earliest main memory algorithm \cite{DBLP:journals/jair/CookH94} developed for substructure discovery. It can perform both supervised and unsupervised substructure discovery. It uses an iterative algorithm to systematically generate larger substructures which are evaluated using Minimum Description Length (MDL~\cite{Rissanen1983AUP}). SUBDUE uses the notion of a \textit{beam} to restrict the number of substructures carried to the next iteration. 

AGM \cite{VLDB/AgrawalS1994} and FSG \cite{ICDM/DeshpandeKK2003} are two popular main memory graph mining algorithms that use the \textit{apriori} concept. These approaches generate frequent ($k$+1) subgraphs from frequent $k$-subgraphs. 
The FSG identifies repeating substructures in graphs using an apriori approach \cite{ICDM/DeshpandeKK2003}. This is different from Subdue since it entails finding interesting substructures in a graph or forest. Canonical labeling is added to the apriori algorithm. The property that identical graphs have identical canonical labeling has been strategically used to identify frequent substructures. FSG determines canonical labels using a flattened graph adjacency matrix. 

Disk-based graph mining algorithms ~\cite{ICDM/PeiHMPCDH2001,book/BunkeS1998,IWWD/AlexakiCKP2001} were developed to deal with larger graph sizes. Portions of the graph are staged into memory for processing using buffer-management techniques. Indexing was used to improve retrieval for staging.  gIndex~\cite{yan2004graph} uses frequent substructures as units for indexing. However, these solutions need explicit data marshalling between disk and main memory entailing developing buffer management strategies. To avoid developing complex buffer management policies, the use of relational databases has been leveraged for graph mining to leverage buffer management and query optimization. 
HDB-Subdue\cite{dawak/PadmanabhanC09} used the RDBMS for graph representation and SQL to discover substructures. This approach was able to scale to handle graphs with millions of edges. However, self-joins on large relations and the number of joins needed for canonical form generation (column sorting instead of row sorting) limited scaling further. 

A number of graph mining methods have shown success in cloud-based deployments\cite{WWW/SuriV2011}. 
Additionally, research has explored patterns in big graphs using the Map/Reduce framework\cite{book/LiuJCMZ2009}. However, the pattern searching technique in \cite{book/LiuJCMZ2009} requires a specified pattern to search for all instances of that pattern in the graph. If we need to find a pattern with the highest compressibility, we cannot supply it beforehand. 
Splitting/partitioning large graphs into manageable chunks for distributed processing is explored in~\cite{sigmod/YangYZK12}.
Scalable substructure discovery on large simple graphs using Map/Reduce has been developed in ~\cite{DaWak:das2015partition}. A graph can be split into partitions in different ways, processed by distributed and parallel architecture, and results from across partitions combined to obtain correct results.

\textit{This paper's focus is on MLNs instead of simple graphs, and specifically HoMLNs. Further, the decoupling-approach is chosen which entails the development of a composition function for its efficient and flexible analysis. It also uses extant substructure discovery for each layer. Correctness of the algorithm is also shown.}

\section{Preliminaries and terminology}
\label{sec:preliminaries}

\begin{wrapfigure}{l}{0.4\columnwidth}
\centering
\vspace{-10pt}
\includegraphics[width=0.45\columnwidth]{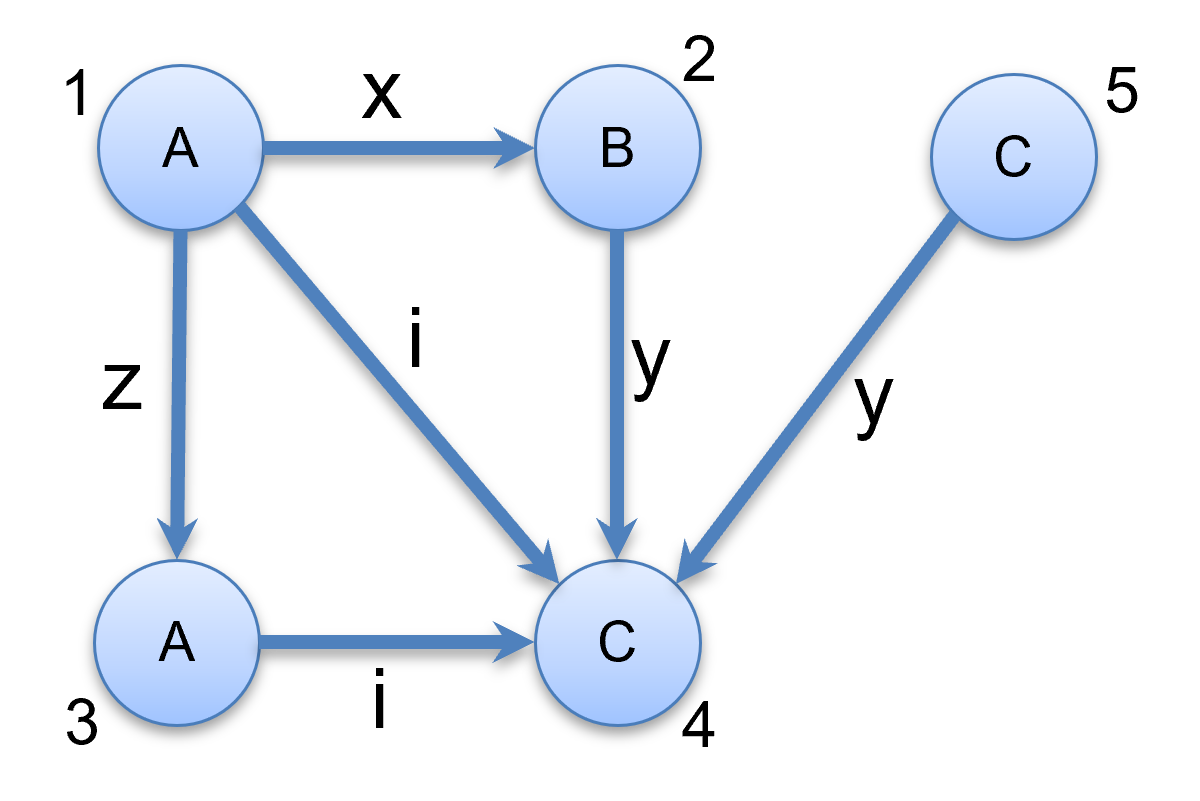}
\caption{\small Input Graph}
\vspace{-10pt}
\label{fig:sample-graph}
\end{wrapfigure}

\noindent{\textbf{Edge List as graph input: }}A graph (layer, inter-layer, or a partition) is represented as a list of unordered edges. Each edge is completely represented by a 5-element tuple\footnote{This representation is generic and can be extended to attributed graphs by using a distinct edge identifier in the edge representation. However, MLN layers are simple graphs by definition.} $<E_l,V_{sid},V_{sl},V_{did},V_{dl}>$
where, 
$E_l$ is edge label, 
$V_{sid}$ is source vertex ID, 
$V_{sl}$ is source vertex label, 
$V_{did}$ is destination vertex ID and 
$V_{dl}$ is destination vertex label.
In this context, it is important to note that the vertex IDs ($V_{sid}$ and $V_{did}$) are guaranteed to be unique. However, vertex labels ($V_{sl}$ and $V_{dl}$) and the edge label ($E_l$) need not  be unique. Table \ref{tab:sample-graph-edge-list} shows the edge list representation for the input graph (or a layer) shown in Figure \ref{fig:sample-graph}.

This 5-element tuple edge representation (that includes direction) is used to represent a $k$-edge substructure (a connected graph with $k$ edges and at most $k+1$ nodes) as a collection of $k$ 1-edge substructures. Our algorithm takes an input graph represented as a text file with a 1-edge substructure per line.

\noindent{\textbf{Adjacency List: }}Adjacency list is a representation where each vertex on which the edges are incident (both in and out) are associated with the node using a list (of 1-edge substructures.) Table ~\ref{table:Adjacency-List} shows the edge list and adjacency list of the graph shown in Figure~\ref{fig:sample-graph}.

\begin{table}[h]
\vspace{-8pt}
\small
  \begin{minipage}{0.15\columnwidth}
    \centering
    \begin{tabular}{|p{1.3cm}|}
      \hline
      \textbf{Edge List} \\
      \hline
      \hline
       (x,1,A,2,B)\\
      \hline
       (z,1,A,3,A)\\
      \hline
       (i,1,A,4,C)\\
      \hline
      (y,2,B,4,C)\\
      \hline
      (i,3,A,4,C)\\
      \hline
      (y,5,c,4,c)\\
      \hline
    \end{tabular}
    \caption{\small Edge List}
    \label{tab:sample-graph-edge-list}
  \end{minipage}\hfill
  \begin{minipage}{0.77\columnwidth}
    \centering
    \begin{tabular}{|p{1.35cm}|p{4.58cm}|}
      \hline
      \textbf{Vertex ID} & \textbf{Adjacency List } \\
      \hline
      \hline
      1 & (x,1,A,2,B), (z,1,A,3,A), (i,1,A,4,C) \\
      \hline
      2 & (x,1,A,2,B), (y,2,B,4,C) \\
      \hline
      3 & (z,1,A,3,A), (i,3,A,4,C) \\
      \hline
      4 & (i,1,A,4,C), (y,2,B,4,C), (i,3,A,4,C), (y,5,c,4,c) \\
      \hline
      5 & (y,5,c,4,c) \\
      \hline
    \end{tabular}
    \caption{\small Adjacency List}
    \label{table:Adjacency-List}
  \end{minipage}
\vspace{-8pt}
\end{table}

\noindent{\textbf{Graph Partitioning:}}
A MLN layer $L_i$ can be partitioned into p partitions ($L_1^i$, $L_2^i$,..., $L_p^i$) for distributed processing. We use
\textit{range-based partitioning}~\cite{DaWak:das2015partition} to create the partitions using the vertex IDs. Each graph partition\footnote{Graph partition and partition are used interchangeably. Adjacency List partition is an adjacency list for a specific graph partition.} is a range of node IDs and the size of each partition need not be the same. There can be missing vertex IDs in a given range. 
As the ranges are disjoint, nodes in adjacency list partitions are also disjoint. Each vertex ID in the range and its adjacency list corresponds to a single adjacency list partition. If neighboring nodes are in two partitions, the edge connecting them will be in the adjacency list of both partitions. As a result, during expansion, 
\begin{wrapfigure}{l}{0.35\columnwidth}
\centering
\vspace{-12pt}
\includegraphics[width=0.4\columnwidth]{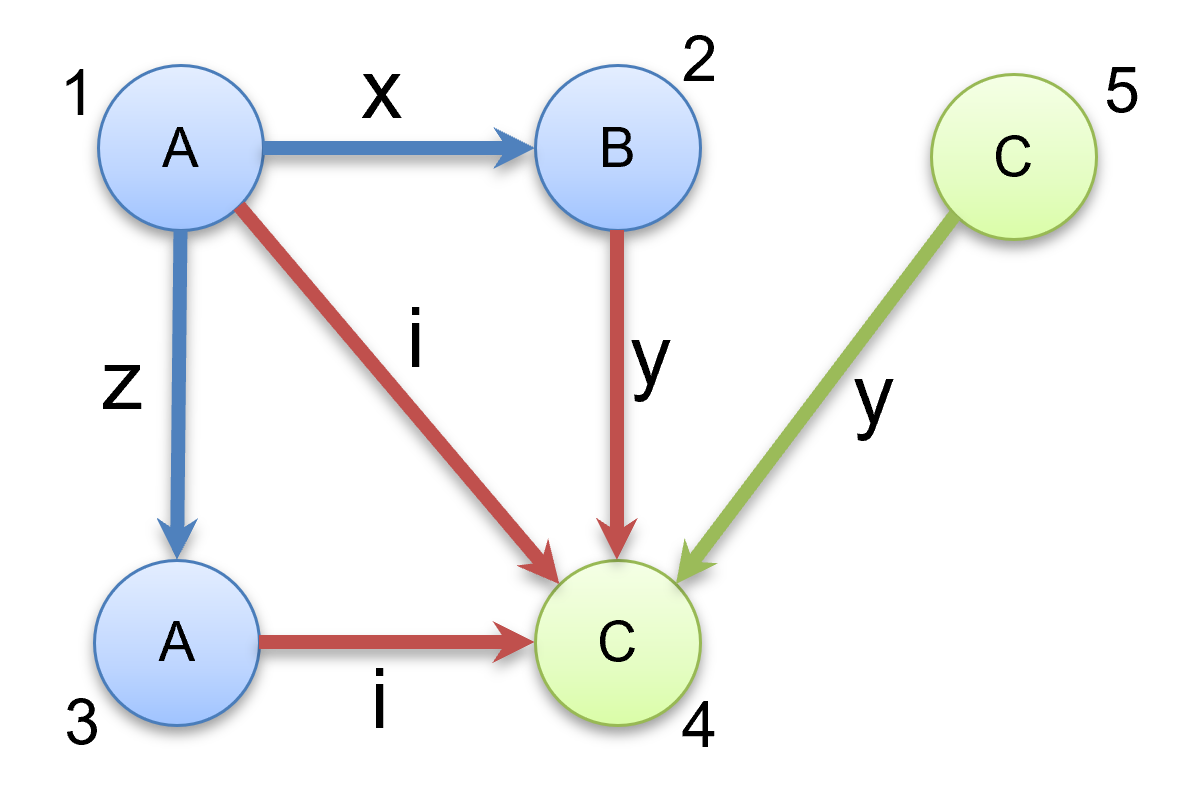}
\caption{\small Partitioned Input Graph}
\vspace{-8pt}
\label{fig:graph-partitions}
\end{wrapfigure}
the same substructure can belong to many graph partitions. As adjacency list partitions are indexed on vertex IDs, each substructure is expanded only once. 

Two partitions of the graph in Figure~\ref{fig:sample-graph} are shown in Figure~\ref{fig:graph-partitions}.
Partition $p1$ is assigned vertex IDs from 1 to 3, whereas partition $p2$ is allotted vertex IDs 4 to 5. Both partitions are connected by red edges, which appear in different adjacency list partitions. The adjacency list partitions are displayed in table \ref{tab:adjacency-list-partitions}. The edges that occur in both adjacency list partitions are $<i,1,A,4,C>$$, $$<y,2,B,4,C>$, and $<i,3,A,4,C>$.

 \begin{table}[h]
 \small
 \centering
  \begin{tabular}{|p{1.4cm}|p{6cm}|}
     \hline
     \textbf{Vertex ID} & \textbf{Adjacency List for Partition P1 } \\
     \hline
      \hline
      1 & (x,1,A,2,B), (z,1,A,3,A), (i,1,A,4,C) \\
      \hline
      2 & (x,1,A,2,B), (y,2,B,4,C) \\
      \hline
      3 & (z,1,A,3,A), (i,3,A,4,C) \\
     \hline
   \end{tabular}
   
   \begin{tabular}{|p{1.4cm}|p{6cm}|}
     \hline
     \hline
       \textbf{Vertex ID} & \textbf{Adjacency List for Partition P2} \\
    \hline
      \hline
      4 & (i,1,A,4,C), (y,2,B,4,C), (i,3,A,4,C), (y,5,c,4,c) \\
      \hline
      5 & (y,5,c,4,c) \\
      \hline
   \end{tabular}
   \caption{Adjacency Lists for Partitions}
   \label{tab:adjacency-list-partitions}
   \vspace{-15pt}
 \end{table}

\begin{wrapfigure}{l}{0.45\columnwidth}
\centering
\vspace{-13pt}
\includegraphics[width=0.5\columnwidth]{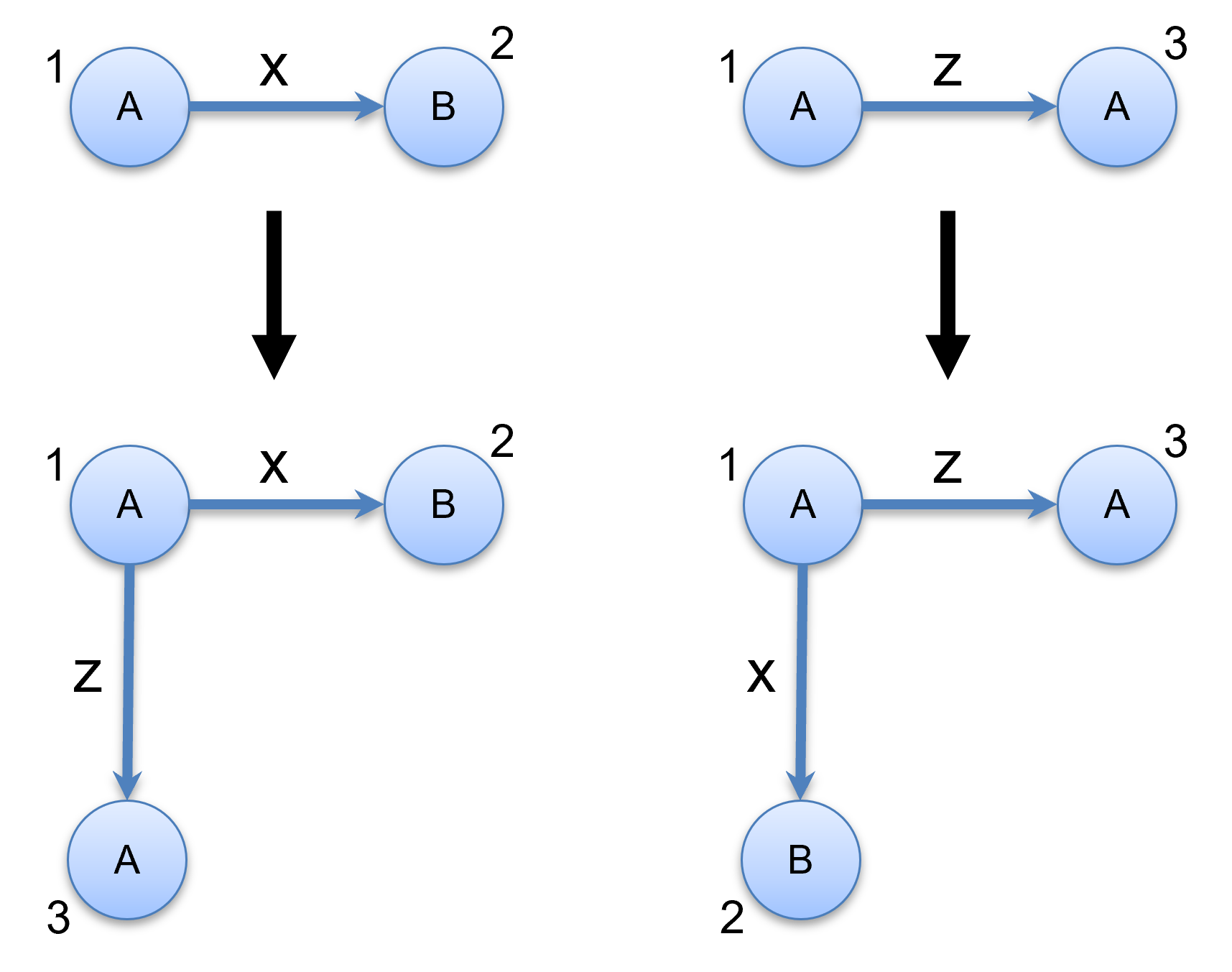}
\caption{\small Duplicate Generation}
\vspace{-12pt}
\label{fig:duplicates}
\end{wrapfigure}

\noindent{\textbf{Independent Expansion of Substructures: }} Substructure discovery uses independent, unconstrained expansion of $k$-edge substructures into $(k+1)$-edge substructures starting from an edge ($k$ = 1 or a 1-edge substructure.)  
Each expansion of each vertex with an edge becomes a separate substructure. This independent unconstrained expansion ensures the generation of all possible substructures in a principled manner thus ensuring completeness of the substructure generation process. However, this independent expansion results in the generation of duplicate substructure instances, which must be removed to prevent incorrect counting of isomorphic substructures (soundness). Figure \ref{fig:duplicates} shows how independent expansion leads to duplicates. If two substructures have the same 
set of vertices (with IDs) and edges (labels) and the same connectivity, they are duplicates. To ensure accurate identification of duplicates and differentiate them from isomorphic substructures, we introduce the concept of a canonical instance and canonical substructure\footnote{A canonical instance has absolute vertex IDs and a canonical substructure has relative vertex IDs.}.

\noindent{\textbf{Canonical Instance for Duplicate Elimination:}} We use lexicographic ordering to convert a substructure instance into its \textbf{canonical representation}. First, we order the edges on the edge label. In case edges have the same edge labels, we order them based on the source vertex label. If source vertex 

\begin{wrapfigure}{l}{0.45\columnwidth}
\centering
\vspace{-12pt}
\includegraphics[width=0.5\columnwidth]{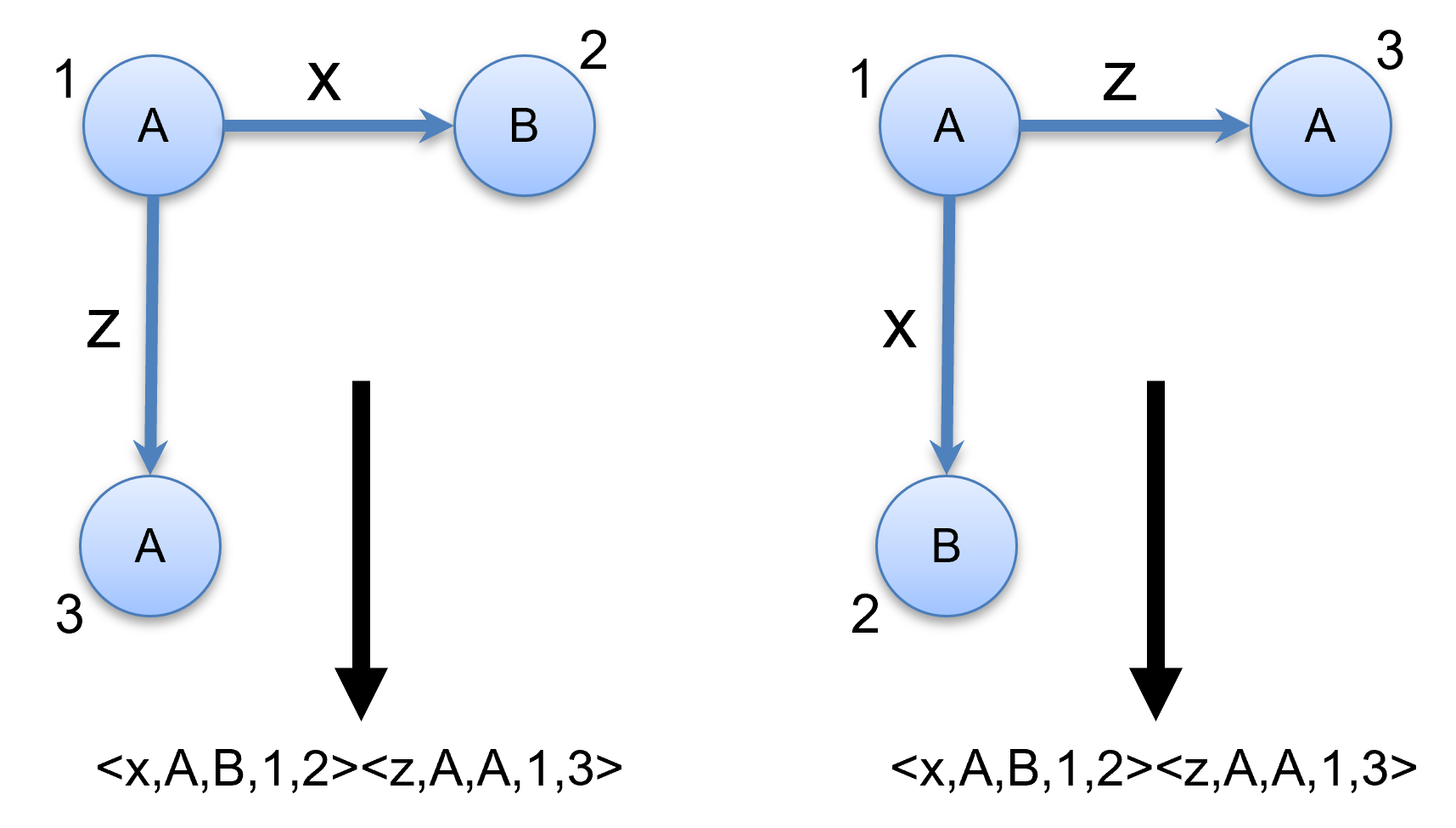}
\caption{\small Canonical Instances Example}
\vspace{-10pt}
\label{fig:Canonical Instances}
\end{wrapfigure}
\noindent labels are also the same, then we further order them based on the destination vertex label. If edge labels and vertex labels are all identical, then we order them based on the source vertex IDs. Finally, if source vertex IDs are also the same, then we order them based on the destination vertex IDs. Using this lexicographic order for \textbf{each 1-edge}, a substructure can be uniquely represented, which is called a canonical $k$-edge instance. If two $k$-edge substructure instances are duplicates, they must have the same ordering of their vertex IDs and therefore, will have the same canonical $k$-edge instance. If we convert the duplicates generated in Figure \ref{fig:duplicates} to canonical instances, they will be the same, as shown in Figure \ref{fig:Canonical Instances}.

The canonical representation of instances is helpful in identifying and removing duplicate instances. However, counting the frequency of isomorphic substructures requires a canonical form that does not rely on absolute vertex IDs. Therefore, we use a canonical form of the substructure that converts vertex IDs to relative vertex IDs.

\noindent{\textbf{Canonical Substructures for identifying Substructure Isomorphs:}} Graph isomorphism refers to the concept of two graphs being structurally identical, with a one-to-one mapping of their vertices, preserving edge connections and edge labels, and edge orientations. Exact matching of substructures is needed to compute frequent substructures accurately. Isomorphs have the same vertex labels and edge labels but different vertex IDs (as they are different instances). To identify them, we need to create a canonical substructure from the canonical instance. For this vertex IDs are changed from absolute to relative ordering. Any two exact substructures will have the same relative ordering of vertex IDs \cite{DaWak:das2015partition}.
This is used to identify and compute the frequency of a substructure. 

\begin{wrapfigure}{l}{0.55\columnwidth}
\centering
\vspace{-10pt}
\includegraphics[width=0.58\columnwidth]{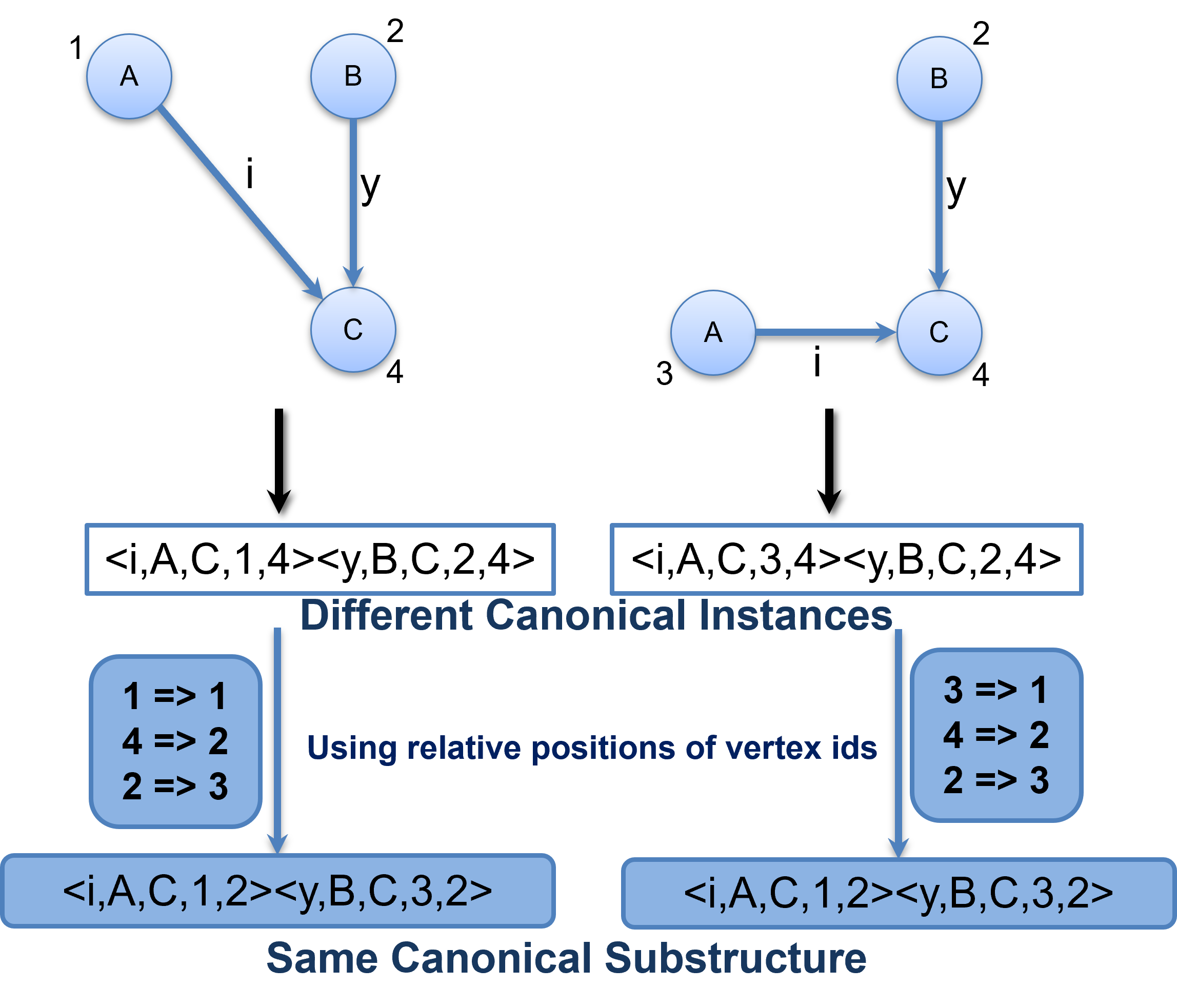}
\caption{\small Substructure Isomorphism Example}
\vspace{-10pt}
\label{fig:Graph Isomorphism}
\end{wrapfigure}

As the canonical instance already follows a lexicographic ordering, we can construct a canonical $k$-edge substructure by ordering the unique vertex IDs in the order of their appearance in the canonical instance. Thus, the canonical substructure can be derived by replacing every vertex ID with its relative position in the instance. The resulting canonical substructure allows us to identify isomorphs easily. Figure \ref{fig:Graph Isomorphism} provides an example of how a canonical substructure is created from the canonical instance. Note that the isomorphs have different canonical instances, but their relative positions following the canonical instance are the same. The relative positioning of vertex ID (1, 4, 2) for the first instance and (3, 4, 2) for the second instance reduces to (1, 2, 3). This leads to both having the exact same canonical substructure.

The process of canonical conversion to both instances and substructures is crucial as it enables us to identify duplicates and isomorphic instances, respectively. 

\subsection{Metrics Used for Ranking Substructures}

\noindent The Minimum Description Length is an information-theoretic, domain-independent metric~\cite{DBLP:journals/jair/CookH94} that has been demonstrated to emphasize the significance of a substructure in terms of its ability to compress a complete graph or forest. Although it is defined, originally, in terms of bits used for graph representation, we use the number of nodes and edges for that purpose. The general formula for MDL (DL(G)/(DL(S) + DL(G$|$S))) represents the description length of the substructure S being evaluated, DL(G$|$S) represents the description length of the graph G when compressed by replacing each instance of the substructure S as a node, and DL(G) (or DL(S)) represents the description length of the original graph (or the substructure S.) The substructure of the graph achieves the highest compression when the MDL value is the highest. Both the frequency of the subgraph and its connectivity have an impact on compression.
The frequency of substructures can also be used as a metric as used in FSG and others.

\section{Architecture And Composition Algorithm}
\label{sec:architecture-algorithm}
\noindent Instead of a single network, we have multiple layers and the goal is to find \textit{interesting} substructures in the MLN (across all layers). Substructures in a Multilayer Network (MLN) can exist in different ways: i) can exist solely in one layer (intra-layer substructure) or ii) can span multiple layers (inter-layer substructure). An algorithm exists for detecting substructures within a graph or a layer. For scalability, this algorithm has been extended to use range-based partitions and generate substructures across partitions correctly\cite{DaWak:das2015partition}. 
Hence, our primary focus is on generating all inter-layer substructures correctly during the composition phase so MDL (or some other metric) can be applied to rank the MLN substructures.

Below we present the \textbf{I}terative \textbf{C}omposition \textbf{A}lgorithm (\textbf{Ho-ICA}) for \textbf{Ho}mogeneous MLNs which performs composition \textit{after each iteration}. That is, substructures for each participating layer are generated during an iteration and the generated layer substructure instances are used for composing \textbf{missing} inter-layer substructures. After that, duplicate elimination is done for generated instances using canonical representation and substructures are counted using canonical substructures to apply MDL. This process repeats after each iteration until the substructure discovery algorithm terminates and outputs the best substructures based on the given parameters.

As inter-layer substructures are composed in every iteration, our conjecture is that this approach should not have any accuracy drop (if done correctly) as all substructures (intra and inter) are generated in each iteration and evaluated using the metric before keeping the \textit{top-beam} substructures for the next iteration. We will validate this conjecture experimentally. We will also provide the operational correctness of the algorithm. Although we may be adding an extra step of computation (composition) as compared to the ground truth computation, yet the efficiency (as compared to ground truth) will improve by an order of magnitude while preserving accuracy, as we are dealing with smaller graphs that can be processed in parallel.

We will experimentally evaluate the developed algorithm for correctness, efficiency, and scalability. We will also discuss how distributed processing resources can be matched to the input graph size to keep the desired response time.

\subsection{Ho-ICA: HoMLN Iterative Composition Algorithm}

\begin{figure}[h]
\centering
\includegraphics[width=\columnwidth]{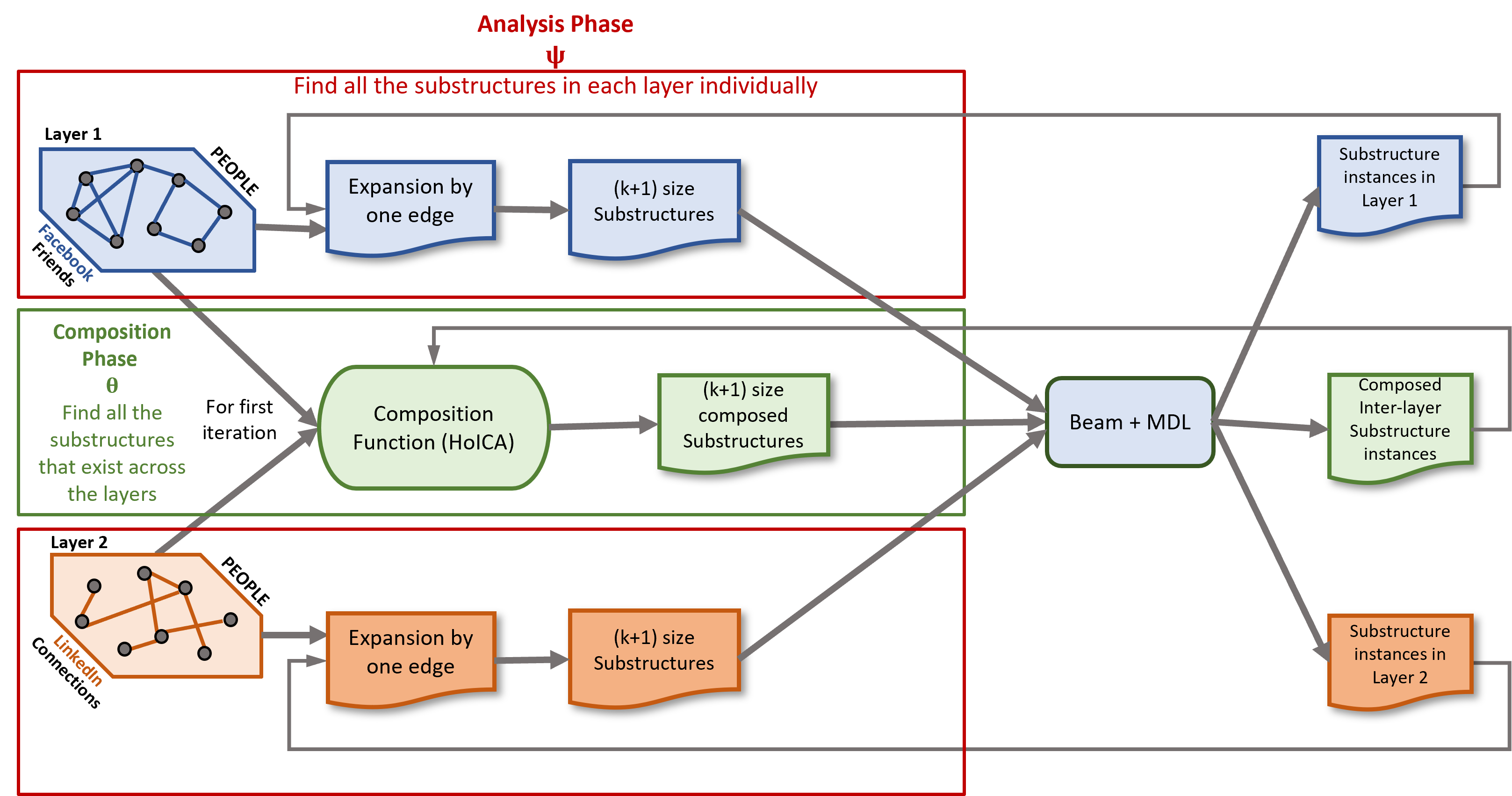}
\caption{\small Decoupling Approach Overview for Iterative Composition}
\label{fig:hoica}
\vspace{-15pt}
\end{figure}

\noindent Figure \ref{fig:hoica} provides a comprehensive diagrammatic overview of Ho-ICA for substructure discovery for a HoMLN. 

This approach generates substructures for each layer in each iteration, converts them to canonical instances, and eliminates duplicates. Composition algorithm is applied at the end of layer substructure generation in each iteration to generate missing inter-layer substructures. After that, all substructures from the individual layers and composed substructures spanning layers (after duplicate elimination) are aggregated. They are converted into canonical substructures for applying a ranking metric (such as MDL) to generate the \textit{top-beam} substructures for the next iteration. Each iteration corresponds to a particular size of substructure. In each iteration, the substructures are expanded by a single edge in their own layer, starting with size one substructures (an edge) for the first iteration. This generates all the intra-layer substructures for that size. Also, in the first iteration which generates 2-edge substructures, the composition function generates inter-layer substructures by expanding instances from each layer using the adjacency list of other layers. In subsequent iterations, composition function will generate composed substructures of size $k+1$, using the composed substructures from the previous iteration (size $k$).

At the end of each iteration, the substructures from each layer and the inter-layer substructures are separated for the next iteration. This is critical for the decoupling principle which stipulates that layers are processed independently and without using the knowledge of the other layers. In the next iteration, substructures from each layer are expanded by an edge from that layer and in the next composition stage, intra-layer substructures are expanded using edge from the other layer and inter-layer substructures are expanded with one edge from either layer and the process repeats. Note that composition can make use of information from both layers.

The input of the Ho-ICA algorithm is a set of composed substructures from the previous iteration, each of size \textit{k-1}, where \textit{k} is at least 2. We then expand these substructures using adjacency lists of other layers, to get inter-layer composed substructures of size \textit{k}. The inter-layer substructures, in addition to the intra-layer substructures generated in the current iteration, are forwarded to the metric evaluation process followed by beam application. Given that the beam substructures now consist of both intra-layer and inter-layer substructures, they are separated (to be consistent with the decoupling philosophy) and sent to their respective tasks for the next iteration. 
Algorithm \ref{alg:ho-ica} describes the composition function used to compose substructures after each iteration.

\begin{algorithm}[h]
\small
\begin{algorithmic}[1]
\REQUIRE
$(S_{k-1})$: Set of 
all substructures of size $k-1$ from previous iteration (composed and layers), where $k$ is at least 2 represented as a sequence of edges $<e_1, \, \ldots, \,e_k>$, where  $e_i$ is 5-element tuple;    $AL^m_{p}  \ldots AL^n_{p}$: Adjacency List partitions for N layers 
\ENSURE
Composed inter-layer substructures of size $k$\;
\FOR {each $(k-1)$ edge instance $I_{k-1} \in (S_{k-1})$} 
        \FOR {each vertex-ID $v \in I_{k-1}$}
            \STATE {$EL_v \leftarrow \{v \in AL^m_{p} \bigcup \ldots \bigcup AL^n_{p} \}$} 
            
            \COMMENT {edge list of $v$ from every layer}
            \FOR {each edge $e \in EL_v$}
                \IF { $e \notin I_{k-1}$}
                \STATE $I^{composed}_k \leftarrow I_{k-1} + e $ 
                
                \COMMENT{Append edge $e$ in lexicographical order}
                \STATE $ S^{composed}_k \leftarrow I^{composed}_k$ 
                
                \COMMENT {Add to composed set $S^{composed}_k$}
                \ENDIF
            \ENDFOR
        \ENDFOR
\ENDFOR
\end{algorithmic}
\caption{Ho-ICA: Iterative Composition Algorithm}
\label{alg:ho-ica}
\end{algorithm}

The algorithm expands each previously generated inter-layer substructure using adjacency lists of all layers (line 3).
\section{Correctness of  Approach}
\label{sec:homln-correctness}

\noindent In this section, we will prove that expanding each substructure instance of a graph or forest one edge at a time followed by duplicate elimination is complete and sound. Completeness means \textit{all} substructures that should to be generated for evaluation are generated without fail. Soundness means that \textit{only} those substructures that should be generated are generated. We need duplicate elimination for soundness. In addition, we will also argue that the composition algorithm presented here is operationally correct and will generate all inter-layer substructures without fail. We will also prove the correctness of the presented Ho-ICA algorithm.

\subsection{Correctness of Independent Substructure Expansion}

Given a forest $F$ of graphs, let  $G(V, E)$ be an arbitrary connected graph\footnote{Without loss of generality, the results extend to all connected components in $F$. As our layers are simple graphs/forest, we are not considering multiple edges between nodes, and loops.} in $F$, where $V$ is the set of nodes and $E$ is the set of edges (directed or undirected). An adjacency list exists for $F$ which is indexed on node ID and its value being the edges incident on that node (directed and undirected). Let $AL_F$ be the adjacency list of $F$.

Let $g$ be a connected substructure instance of $G$ with $k$ edges and at most $k+1$ nodes\footnote{It is easy to see that the first iteration on an edge using the adjacency list generates all 2 edge substructures containing that edge.}. Let $n'$ be a node in $g$ and $d'$ be its degree in $g$. Let the degree of $n'$ in $G$ be $d$. \textit{Independent expansion refers to each substructure instance of size $k$ (i.e., $k$ edges) being expanded without considering any other substructure of size $k$.}

\begin{lemma}
\label{lem:instance-completeness}
\textit{Independent expansion of the substructure instance $g$ generates all instances that should be generated from $g$.}
\end{lemma}

\begin{IEEEproof}
During independent expansion, an arbitrary node $n'$ of $g$ is expanded by one edge (along with the connected node) that is not already present in the substructure instance, but is incident on $n'$ in the graph $G$. This edge is added to the subgraph instance $g$ increasing the number of edges (and may be nodes) by 1. It also increases the degree of the node $n'$ by 1 and increases the size of the substructure by 1. This becomes a new substructure instance with at most $k+2$ nodes and $k+1$ edges.
This is done for each edge incident on $n'$ in $G$ that is not in $g$. For node $n'$, the number of such 1-edge expansions will be $(d-d')$. This process is repeated for all nodes in $g$ and \textit{separate} substructure instances are generated for each expansion. The total number of substructure instances generated for each substructure instance $g$ depends on the degree of individual nodes in $g$ and the degree of corresponding nodes in $G$. This is repeated for all substructure instances of size $k$ in iteration $k$ generating all substructure instances of size $(k+1)$ in $k^{th}$ iteration. 
Therefore, as each node in $g$ is expanded to the fullest extent, this process will generate all possible substructure instances of size $(k+1)$ of graph $G$ in iteration $k$, that should be generated from $g$.
\end{IEEEproof}

\begin{lemma}
\label{lem:completeness}
\textit{In the absence of any constraints, the application of Lemma~\ref{lem:instance-completeness} repeatedly on substructure instances starting with the substructures of size 1 achieves completeness.}
\end{lemma}

\begin{IEEEproof}
The above process starts with 1-edge substructure instances (essentially all edges of graph $G$.). In iteration 1 all substructure instances of size 2 are generated using the above procedure. This process is repeated on increasing substructure instances until no more expansion can be performed. The final substructure instance generated with unconstrained expansion is $G$ itself. This expansion until no more instances can be generated achieves completeness.
\end{IEEEproof}

Note that independent expansion of a single substructure instance only generates distinct instances. However, when two different substructure instances are expanded independently, duplicate instances may be generated. An example of this has been shown in Figure~\ref{fig:duplicates}. Instances at the bottom were generated by expanding on different substructure instances at the top by an edge. The resulting substructures become identical in node IDs and labels as well as edge labels as can be seen after the canonical ordering. Only one needs to be kept and the other eliminated for soundness. We have indicated earlier how canonical form of substructures are used for identifying duplicate instances for elimination.

\begin{lemma}
\label{lem:duplicate-elimination}
\textit{Independent expansion of instances of $G$ with duplicate elimination at the end of each iteration achieves soundness.}
\end{lemma}

\begin{IEEEproof}
Consider all subgraph instances of size $k$. When these are expanded independently in the $k^{th}$ iteration, a large number of instances of size $(k+1)$ are generated. As indicated above, there may be duplicate instances among them. By converting each substructure to its canonical form and sorting them, duplicates are brought together. Duplicates are eliminated by keeping only one instance. This ensures soundness.
\end{IEEEproof}

\begin{figure*}
    \centering
    \includegraphics[width=2\columnwidth]{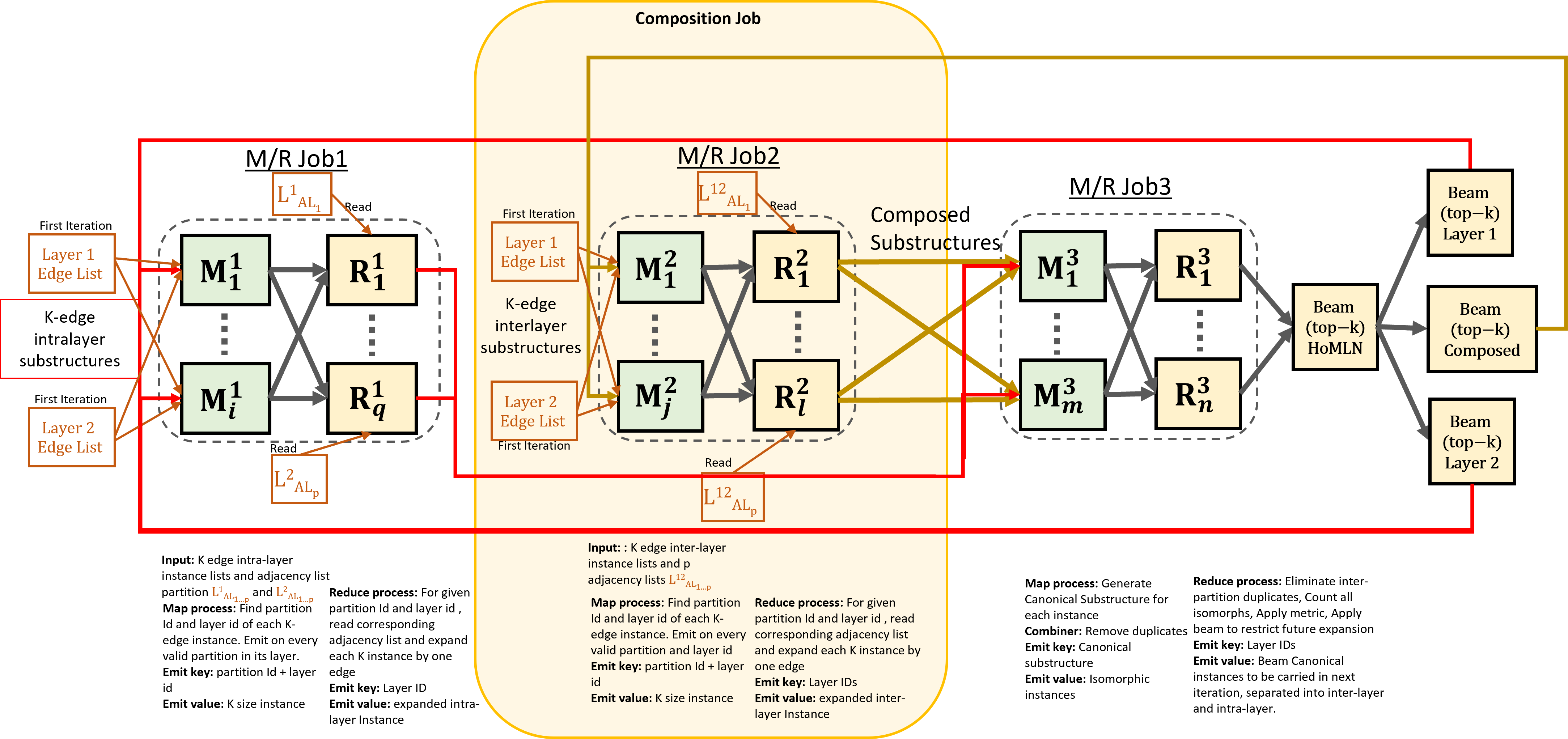}
    \caption{Ho-ICA Architecture using Map/Reduce}
    \label{fig:MR-hoica}
    \vspace{-17pt}
\end{figure*}

\begin{theorem}
\label{thm:sound-complete}
 \textit{Independent expansion of instances followed by duplicate elimination in each iteration is complete and sound.}   
\end{theorem}

\begin{IEEEproof}
Follows directly from Lemmas~\ref{lem:completeness} and~\ref{lem:duplicate-elimination}.
\end{IEEEproof}

\subsection{Correctness of the Ho-ICA Algorithm}

\noindent Here we need to show that Ho-ICA algorithm generated \textbf{all} inter-layer instances without fail. Note that intra-layer instances are correctly generated, as established by Theorem~\ref{thm:sound-complete}.

We will use proof by induction. We will show that for any iteration \textit{k}, Ho-ICA correctly generates all results for the next iteration \textit{(k + 1)}. Following is the proof:

\begin{lemma}
\label{lem:ho-ica-correctness}
\textit{For any iteration $i$, the Ho-ICA algorithm correctly generates all inter-layer instances.}
\end{lemma}

\begin{IEEEproof}
\noindent \textbf{Base Case:}
For iteration 1, Ho-ICA uses size 1 substructures (i.e., the edge list) of each layer and generates size 2  inter-layer substructure instances using the adjacency lists of the other layers. As it utilizes all edges in each layer to pair with edges from any other layer, the Ho-ICA will generate every size 2 inter-layer instance. With duplicate elimination all size 2 distinct inter-layer substructures are generated.
Thus, Ho-ICA correctly generates all size 2 substructure instances in the first iteration correctly.

\noindent \textbf{Inductive Case:} 
We assume that for some arbitrary iteration \textit{k-1}, the algorithm has correctly generated all inter-layer substructure instances of size $k$.

In iteration $k$, Ho-ICA will expand inter-layer substructure instances from the previous iteration (iteration \textit{k-1}) to generate expanded inter-layer substructure instances. Specifically, it employs the appropriate adjacency list for each layer to which the to be composed instance belongs. Through an unconstrained expansion on every vertex ID of the instance using these adjacency lists, Ho-ICA ensures the generation of all possible expanded inter-layer instances. For two layers $L_1$ and $L_2$,
\begin{enumerate}
    \item all intra-layer substructures are generated by expanding $L_1$ substructures using $L_2$ adjacency list and $L_2$ substructures using $L_1$ adjacency list.
    \item inter-layer substructures are generated using composed substructures from the previous iteration and expanding them using both $L_1$ and $L_2$ adjacency lists.
\end{enumerate}

After the unconstrained expansion, duplicate elimination is done to ensure soundness. Thus, Ho-ICA accurately generates all instances for the iteration \(k\) using all instances from iteration \(k-1\).
\end{IEEEproof}
\section{Implementation Using Map/Reduce}
\label{sec:implementation}

\noindent In this section, we discuss the implementation of the decoupling-based iterative substructure discovery algorithm (Ho-ICA) discussed in Section~\ref{sec:architecture-algorithm}. We use the Map/Reduce framework to leverage distributed and parallel processing. Both range partitioning and substructure discovery have been implemented using the Map/Reduce framework.

The iterative composition algorithm for substructure discovery in HoMLNs follows the two distinct phases of the decoupling approach, namely, analysis and composition phases. Figure \ref{fig:MR-hoica} shows the overall flow of Ho-ICA using Map/Reduce. 

The algorithm comprises three Map/Reduce jobs for each iteration. The first pair generates layer substructure instances using independent and unconstrained expansion. The third pair counts substructures, applies MDL, and generates \textit{top-beam} substructures for the next iteration. The composition function Ho-ICA is encapsulated in the second Map/Reduce pair. The composition function will now expand inter-layer composed instances by using the adjacency list as explained in Algo.~\ref{alg:ho-ica}. As shown, only iteration 1 is different from the rest.
\section{Experimental Evaluation and Validation}
\label{sec:experiments-validation}

\noindent Experiments were conducted on the Expanse cluster at SDSC (San Diego Supercomputer Center), details shown in Table~\ref{table:expanse}. 

\begin{table}[h]
\small
\centering
\begin{tabular}{|p{0.25\columnwidth}|p{0.6\columnwidth}|}\hline
 \textbf{Compute Node Component}&\textbf{Configuration}\\\hline
\hline
Node Count & 728 \\
\hline
Cores/Node & 128 built on 2 processors (64 cores each)\\
\hline
Processor & AMD EPYC 7742\\
\hline
Memory &  256 GB DDR4 DRAM\\
\hline
Storage & 1TB Intel P4510 NVMe PCIe SSD \\
\hline

\end{tabular}
\caption{\small Expanse System Details}
\label{table:expanse}
\end{table}

\noindent {\textbf{Data Set Generation: }}
In the context of our substructure discovery algorithm, when we model a data set as a HoMLN, each layer is considered as an independent graph. Consequently, we can transform a large simple graph into a MLN by partitioning it into layers. This approach provides the flexibility to assess the algorithm's performance across MLNs of varying sizes while maintaining control over other graph characteristics, such as the edge distribution among layers. Furthermore, for empirical validation, we can embed substructures with known frequencies to ensure consistent identification of substructures and their frequency whether the data set is processed as a single graph or as a MLN. Hence, our methodology involves generating substantial single graphs with embedded substructures, followed by the creation of layers for the HoMLN.

\begin{table}[h]
\vspace{-5pt}
\small
\renewcommand{\arraystretch}{1}
\centering
\begin{tabular}{|p{.17\columnwidth}|p{.23\columnwidth}|p{.08\columnwidth}|p{.3\columnwidth}|} \hline        
 \textbf{Data Set}& \textbf{Used For} & \textbf{Beam Value} & \textbf{M/R configs (per layer)} \\ \hline 
 Synthetic& Accuracy, Response Time& 4,8,12 & 2M/2R, 4M/4R, 8M/8R\\
 \hline 
 Synthetic Large& Response Time, Scalability & 4 & 4M/4R, 8M/8R, 16M/16R, 32M/32R\\
 \hline \hline 
 Amazon& Response Time, Scalability & 4 & 8M/8R, 16M/16R, 32M/32R, 64M/64R \\
 \hline 
 LiveJournal & Response Time, Scalability & 4 & 8M/8R, 16M/16R, 32M/32R, 64M/64R\\
 \hline
        \end{tabular}
\caption{Data Set Description}
\label{tab:all-datasets}
\vspace{-12pt}
\end{table}

\begin{table}[h]
\small
\centering
\begin{tabular}{|p{2cm}|p{0.5cm}|p{0.9cm}|p{1.3cm}|p{1.3cm}|} \hline        
\textbf{Base Graph \#nodes, \#edges} & \textbf{GID} & \textbf{Edge  Dist.} & \textbf{\#L1 edges} & \textbf{\#L2 edges} \\ \hline 
\multirow{3}{*}{50KV, 100KE} & 1 & 50/50 & 49797 & 50203 \\ 
\cline{2-5} 
& 2 & 70/30 & 69862 & 30138 \\
\cline{2-5}
& 3 & 90/10 & 89899 & 10101 \\ \hline 
\multirow{3}{*}{100KV, 500KE} & 4 & 50/50 & 249953 & 250047 \\
\cline{2-5}
& 5  & 70/30 & 349722 & 150278 \\
\cline{2-5}
& 6  & 90/10 & 449691 & 50309 \\ \hline 
\multirow{3}{*}{400KV, 1000KE} & 7 & 50/50 & 500006 & 499994 \\
\cline{2-5}
& 8  & 70/30 & 700077 & 299923 \\
\cline{2-5}
& 9  & 90/10 & 899994 & 1000006 \\ \hline 
1000KV, & 10 & 50/50 & 1998995 & 2001005 \\
\cline{2-5}
4000KE& 11  & 70/30 & 2799577 & 1200423 \\
\cline{2-5}
& 12  & 90/10 & 3599822 & 400178 \\ \hline 
Amazon & 13 & 50/50 &  1305431&  1305160\\
\cline{2-5}
.74MV, 2.6ME & 14  & 70/30 &  1827821&  782770\\
\cline{2-5}
& 15  & 90/10 &  2349423&  261168\\ \hline 
LiveJournal & 16 & 50/50 &  34489570&  34504203\\
\cline{2-5}
4.9MV, 69ME & 17  & 70/30 &  48295818&  20697955\\
\cline{2-5}
& 18  & 90/10 &  62096307&  6897466\\ \hline

\end{tabular}
\caption{Data Sets with Edge Distributions}
\label{tab:layers-datasets}
\vspace{-5pt}
\end{table}

\noindent {\textbf{Data Set Description: }}
We conducted experiments on real-world and synthetic data sets to evaluate the effectiveness, correctness, speedup concerning different configurations of mappers and reducers, and scalability of our approach. Table \ref{tab:all-datasets} outlines the data sets used. We have utilized synthetic graphs of various sizes with multiple embedded substructures of known frequency. This helps us to verify our results. Additionally, real-world data sets from Amazon~\cite{BRSLLP,BoVWFI} and LiveJournal~\cite{LiveJournal} were also used for our experiments.

For these data sets, diverse two-layer HoMLNs were generated with three edge distributions (50/50, 70/30, and 90/10 \underline{choosing edges randomly}) for testing, details in Table~\ref{tab:layers-datasets}. 

\subsection{Empirical Correctness}

\noindent To verify the correctness of our algorithms, we conducted tests on small synthetic graphs generated by Subgen with sizes ranging from 50KV/100KE to 1MV/4ME with predefined embedded substructures. \textbf{Subdue \cite{DBLP:journals/jair/CookH94} and Ho-ICA when run on these graphs, consistently discovered the same substructures and frequency.}

\subsection{Accuracy}

\noindent As indicated above, Ho-ICA consistently exhibited full accuracy, a validation further supported by our experimental analyses. To verify the correctness on large graphs, we embedded substructures of known size and frequency to test the correctness of the proposed approach.  
Across various synthetic data sets, Ho-ICA also consistently identified all instances of the embedded substructure correctly with exact frequencies.

\subsubsection{Effect of Layer Distribution on Accuracy}

Ho-ICA consistently attains full accuracy regardless of the layer distribution, displaying resilience to variations in the distribution ratio — be it 50/50, 70/30, or 90/10.

\subsubsection{Effect of Beam on Accuracy}

Ho-ICA is able to consistently achieve full accuracy at a beam size of 4 (default value). Thus, increasing beam size further will not have any effect on accuracy except more computation is involved.

\subsubsection{Effect of Layer Connectivity on Accuracy}

This experiment was to make sure Ho-ICA correctly identifies interesting substructures whether the layer graph is connected or not. Again, Ho-ICA displayed full accuracy even when the layer graph was disconnected. This is consistent with the correctness proof in Section~\ref{sec:homln-correctness}.

\subsection{Response Time and Scalability: Synthetic Graphs}

\noindent To validate speedup using distributed processing, we varied the number of partitions of the same graph along with increasing the number of mappers and reducers to maximize parallelization. The objective was to observe how response time varies when resources are increased. This will also indicate the speedup we can achieve (linear trend or exhibits diminishing returns). Large synthetic and real-world data sets were utilized to verify the speedup and scalability of our approach.

\subsubsection{Synthetic Graphs}
To analyze the speedup achieved on synthetic data sets, various data set sizes were explored - 50K vertices and 100K edges (50KV100KV), 100K vertices and 500K edges (100KV500KE), 400K vertices and 1 million edges (400KV1ME), and 1 million vertices and 4 million edges (1MV4ME). Regardless of the layer size, each layer was partitioned into the same number of partitions. We experimented with 8, 16, 32, and 64 partitions, maintaining an equal number of mappers and reducers. We saw a similar trend in each data set. Here, we have focused on the largest data set, 1MV4ME. Results for all data sets are shown in Fig. \ref{fig:speed-rlw_syn} (b).

\begin{figure}[h]
\vspace{-15pt}
    \centering
\includegraphics[width=0.8\columnwidth]{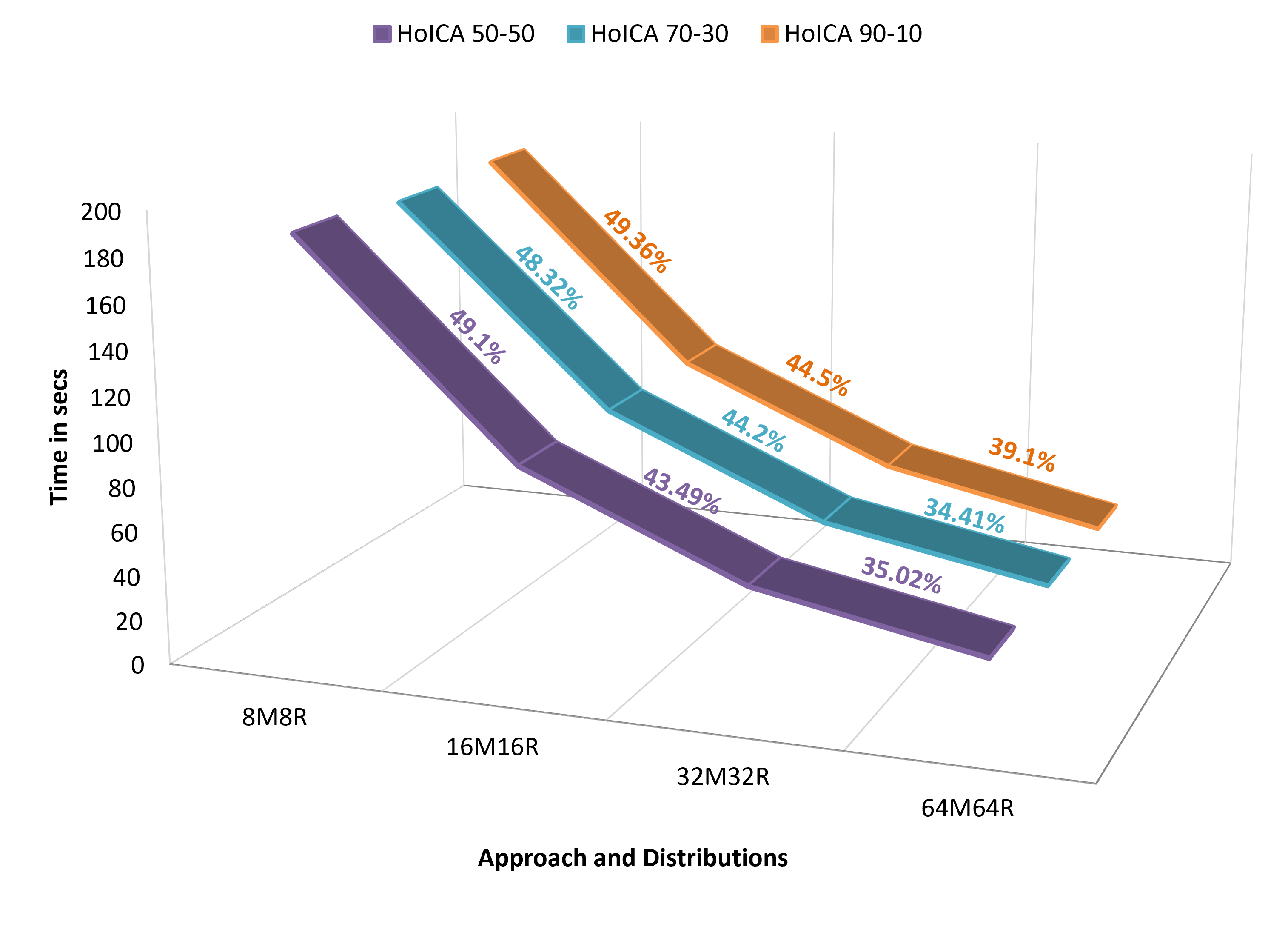}
    \vspace{-15pt}
    \caption{\small Speedup Analysis: 1MV4ME Data Set}
    \label{fig:speed-sca-1m4m}
    \vspace{-10pt}
\end{figure}

For Ho-ICA, Fig. \ref{fig:speed-sca-1m4m} reveals a speedup of at least \textbf{48.32\%} as we increased the number of partitions and the numbers of mappers/reducers from 8 to 16. Subsequently, a speedup of at least \textbf{43.49\%} was observed when further increasing them from 16 to 32. Finally, a speedup of at least \textbf{34.41\%} was noticed when further increasing them from 32 to 64. However, the speedup achieved was not linear; doubling the mappers and reducers does not halve the time taken. As the number of partitions increases, the overhead of writing more files and shuffling comes into effect.

This observed speedup can be attributed to the more partitions of the graphs, processed by an increased number of processors in parallel. Smaller-sized partitions contribute to a reduced computational load on each processor, leading to higher speedup. However, the diminishing speedup suggests that increasing the number of partitions does not continuously improve performance. Higher number of partitions allows more instances to belong to multiple partitions, increasing the potential for inter-partition duplication. This leads to redundant work in different processors. Moreover, excessively small input splits can negatively impact Hadoop performance, introducing overhead related to task setup, communication, and coordination. Conversely, overly large input splits can result in an uneven distribution of processing workload among nodes, leading to inefficient resource utilization. Hence, determining the optimal number of input splits is crucial for optimizing Hadoop job performance. If the objective is to maximize efficiency, employing 16 Mappers/Reducers appears to be the most effective choice as it provides the best speedup. However, in case there is surplus processing power available, 64 Mappers/Reducers still yield a respectable speedup of around 30\% while taking the least amount of time.

\subsubsection{Effect of Layer Distribution}

As evident from Figure \ref{fig:speed-sca-1m4m}, layer distribution does not significantly impact the efficiency of Ho-ICA. This behavior stems from the algorithm's implementation, where all layers are processed within the same Map/Reduce job, enabling effective load balancing and ensuring uniform load distribution among processors.

\subsection{Response Time and Scalability: Real-World Graphs}

\noindent We have also employed real-world graphs to demonstrate scalability and speedup as the characteristics of a real-world graph may be quite different from that of a synthetically generated graph. Two real-world data sets - Amazon and LiveJournal - were used with sizes of 0.74MV2.6ME and 4.9MV69ME, respectively. These data sets were partitioned into 8, 16, 32, and 64 partitions, and experiments were performed using same number of mappers and reducers for each.

\begin{figure}[h]
\vspace{-12pt}
    \centering
\includegraphics[width=0.8\columnwidth]{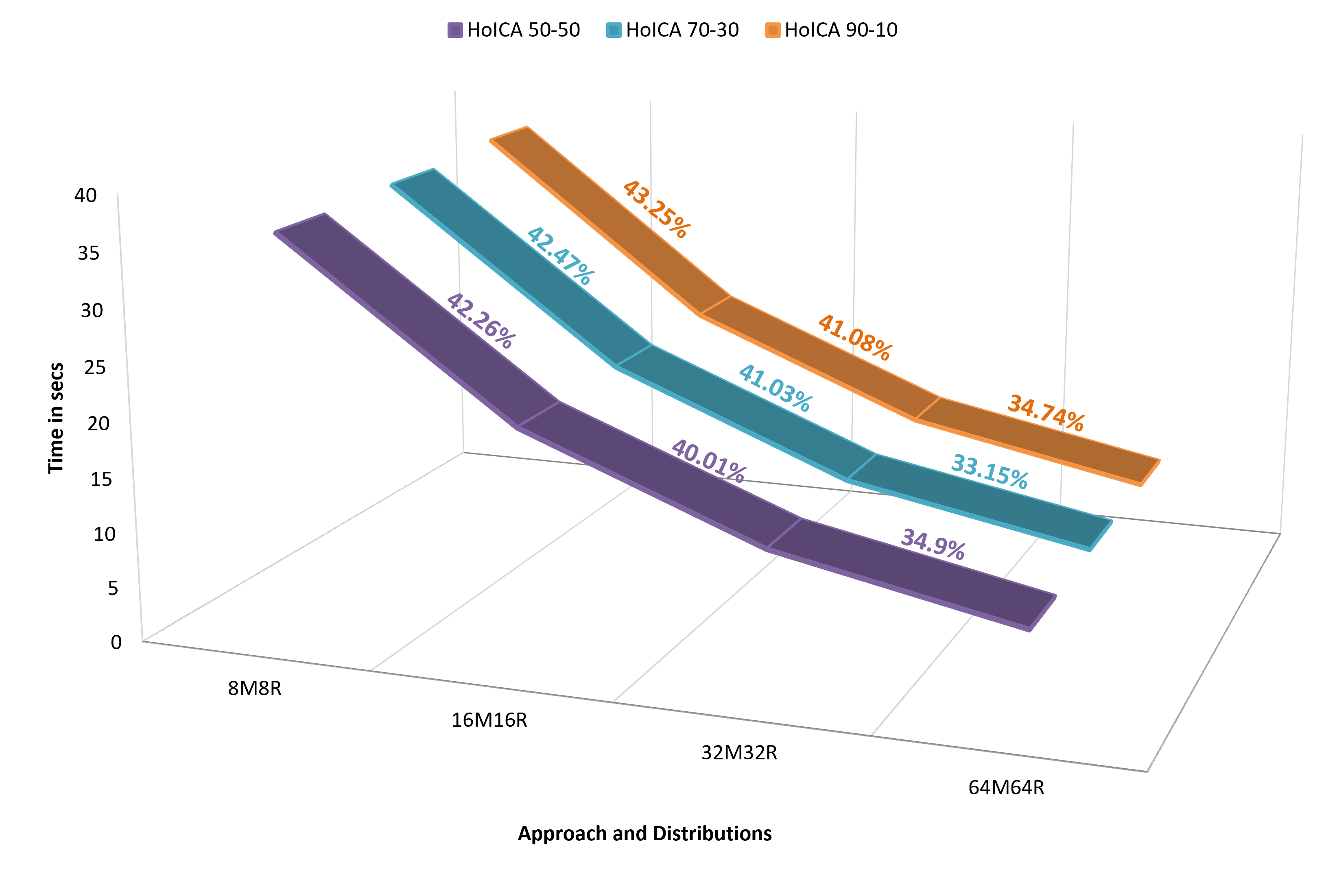}
    \vspace{-10pt}
    \caption{\small Speedup Analysis: Amazon Data Set}
    \label{fig:speed-amazon}
    \vspace{-8pt}
\end{figure}

\noindent \textbf{Amazon (0.74MV2.6ME)}: The speed in Figure~\ref{fig:speed-amazon}, is very similar to the speed up seen for synthetic graphs. Across distributions, we observe speedups of at least \textbf{42.26\%}, \textbf{40.01\%} and \textbf{33.15\%} going from 8 to 16, 16 to 32, and 32 to 64 mappers/reducers, respectively.

\begin{figure}[h]
\vspace{-12pt}
    \centering
    \includegraphics[width=0.75\columnwidth]{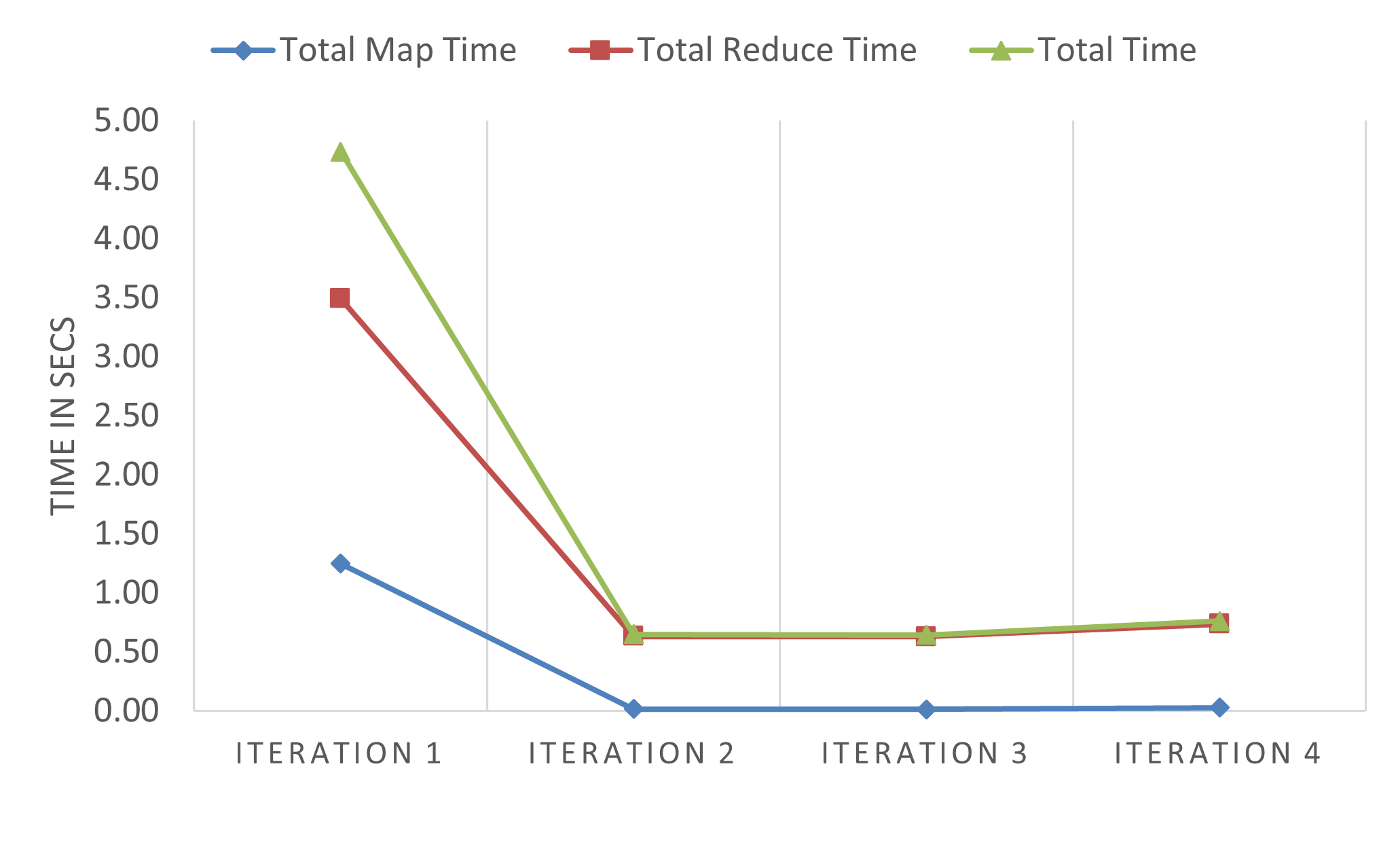}
    \vspace{-14pt}
    \caption{\small Composition Job for Amazon Data Set}
    \label{fig:composition-job-ica-amazon}
    \vspace{-10pt}
\end{figure}

As the focus of this paper is on the composition algorithm, its correctness, and implementation using Map/Reduce, Figure \ref{fig:composition-job-ica-amazon} shows only the composition time for the Amazon data set. Note that there were no embedded substructures as this is a real-world data set. From Figure \ref{fig:composition-job-ica-amazon}, it can be seen that composition takes most time in iteration 1 and is almost flat for other iterations. This is because, in iteration 1, \textbf{all} edges are composed whereas in subsequent iterations, only \textit{top-beam} substructures are composed.

\begin{figure}[h]
\vspace{-10pt}
    \centering
    \includegraphics[width=0.75\columnwidth]{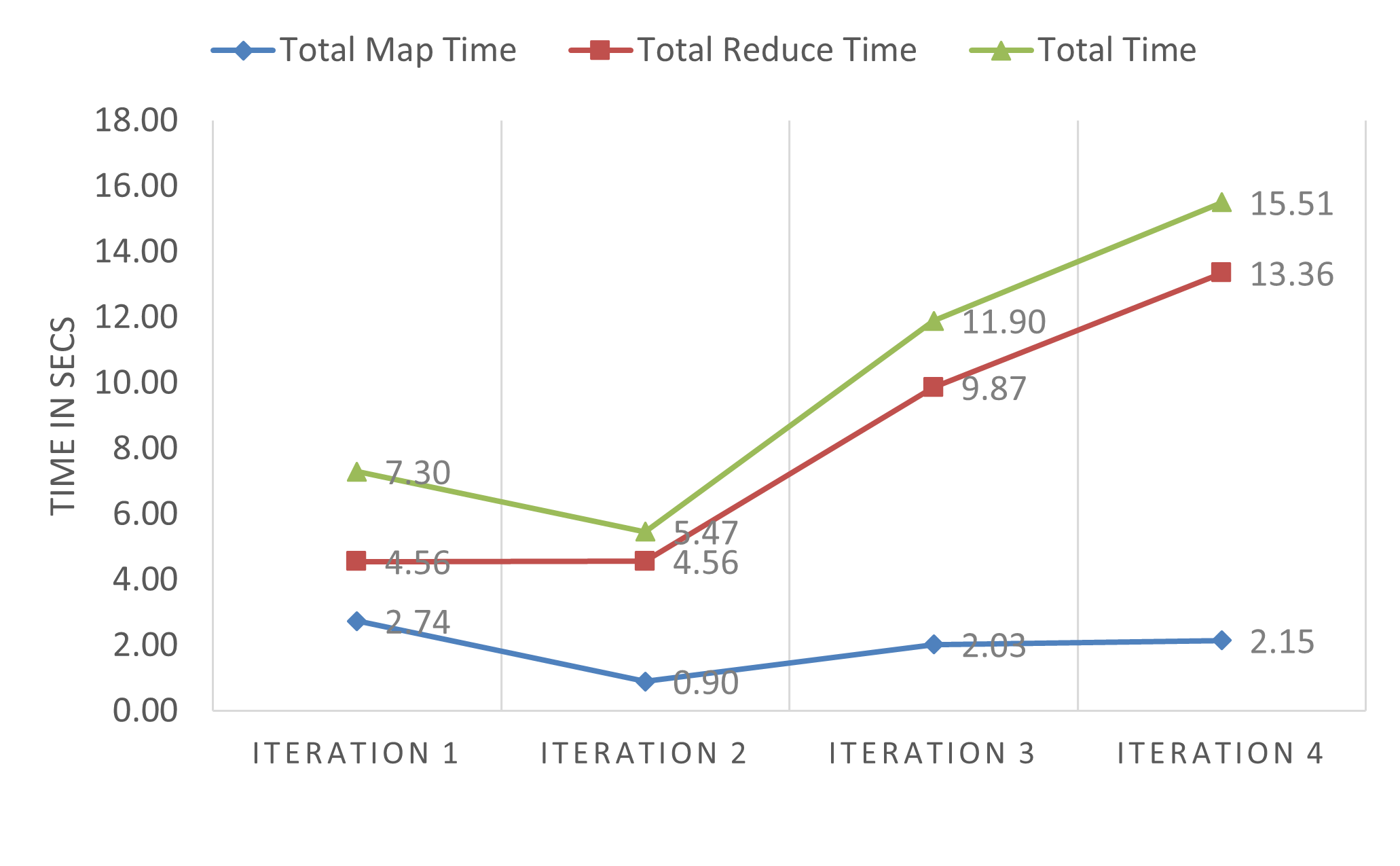}
\vspace{-14pt}
    \caption{\small Composition Job for 1MV4ME Data Set}
    \label{fig:composition-job-ica-1m4m}
    \vspace{-10pt}
\end{figure}

On the other hand, in Figure \ref{fig:composition-job-ica-1m4m} the same composition time is displayed for the synthetic data set 1MV4ME. As this data set has an embedded substructure of size 5, iterations 3 and 4 see an increase in composition time as compared to iterations 1 and 2. Larger substructures (in higher iterations) have more vertex IDs that they can be expanded on, requiring more computational work. This leads to a steady increase in the reduce time from iteration 2 onwards as expansion is done in the reducer. The map time only slightly increases, as the mapper solely handles the routing of substructures based on vertex IDs. This implies that the map time is solely dependent on the number of substructure instances received by the mapper, irrespective of their size. It is essential to note that the dip from iteration 1 to iteration 2 is due to the beam being applied after the first iteration. In other words, all the edges in the graph are processed in the first iteration, but only the top beam substructures are processed starting from the second iteration. This greatly reduces the number of substructures considered for expansion. Thus, a significant dip is observed between iteration 1 and iteration 2.

\noindent \textbf{LiveJournal (4.9MV69ME)}: Here, trends are similar to those seen in the Amazon data set.
This data set is \textbf{notably larger at 4.9 Million Vertices and 69 Million Edges} than the previous data sets. The average speedup while varying the number of partitions and mappers/reducers is greater than that observed in the Amazon data set.

Specifically, a consistent decrease of nearly \textbf{50\%} is observed in transitions from 8 to 16 and 16 to 32 partitions. However, from 32 to 64 partitions, the speedup diminishes to 39\%, but it remains higher compared to other data sets. This can be attributed to the larger size of the graph, which helps alleviate the added overhead from adding more tasks/processors. Having a large overhead due to a higher number of tasks for a small data set would diminish the speedup, but for a sufficiently large data set like LiveJournal, doubling the processors results in nearly double speedup (\textit{approx. 50\% reduction in time}).

\subsection{Response Time for All Experiments Performed}

\begin{figure}[h]
    \centering
\includegraphics[width=1\columnwidth]{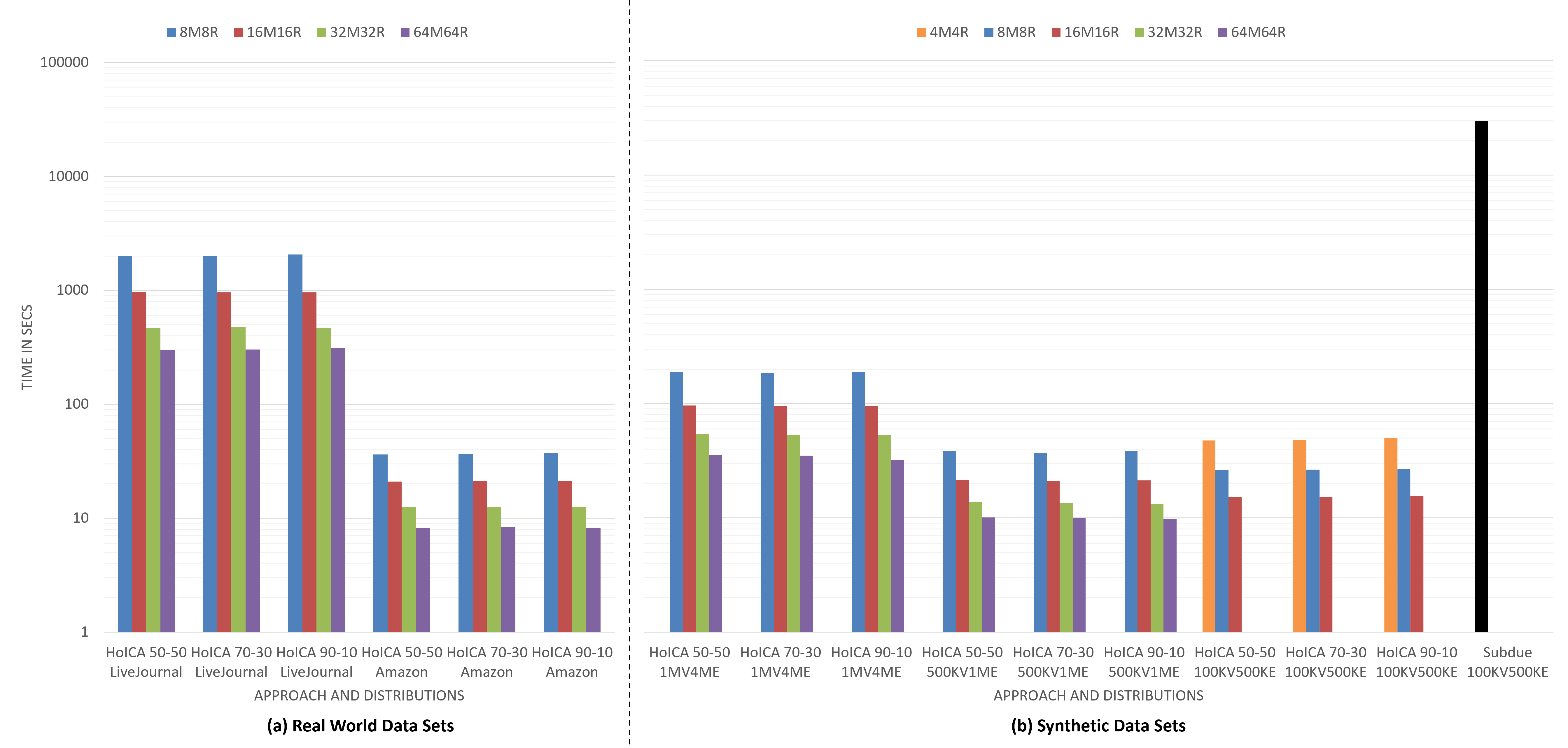}
    \caption{Comprehensive SpeedUp Experiment Results}
    \label{fig:speed-rlw_syn}
\end{figure}

\noindent Figure \ref{fig:speed-rlw_syn} shows all the experiments that were performed for all the data sets shown in Table~\ref{tab:layers-datasets}. 

\vspace{-5pt}
\section{Conclusions and Future Work}
\label{sec:conclusion}

\noindent This paper introduces a scalable algorithm for substructure discovery for  HoMLNs, using a decoupling-based strategy. A composition algorithm was presented and implemented using Map/Reduce, demonstrating its correctness and scalability. 

Future work includes further improving efficiency by skipping composing after each iteration, still preserving accuracy. This seems to be a challenge as it is a trade-off among efficiency, accuracy, and graph characteristics. 

\bibliographystyle{IEEEtran} 

\small{
\bibliography{IEEEabrv,./bibliography/arsh,./bibliography/kiranbolaj,./bibliography/itlabPublications,./bibliography/itlabTheses,./bibliography/itlabCollections-1,./bibliography/itlabCollections-2}
}

\end{document}